\documentstyle[aps,prl,twocolumn,epsf,floats]{revtex}
\draft
\begin{document}
\wideabs{

\title{Fractionally charged magneto-excitons}

\author{
   Arkadiusz W\'ojs}
\address{
   Department of Physics, 
   University of Tennessee, Knoxville, Tennessee 37996 \\
   Institute of Physics, 
   Wroclaw University of Technology, Wroclaw 50-370, Poland}

\author{
   John J. Quinn}
\address{
   Department of Physics, 
   University of Tennessee, Knoxville, Tennessee 37996}

\maketitle

\begin{abstract}
The photoluminescence (PL) spectrum of a two-dimensional electron 
gas (2DEG) in the fractional quantum Hall regime is studied.
The response of the 2DEG to an optically injected valence hole 
depends on the separation $d$ between the electron and hole layers.
At $d$ smaller than the magnetic length $\lambda$, the PL spectrum 
shows recombination of neutral ($X$) and charged ($X^-$) excitons.
At $d>\lambda$, the hole binds one or two Laughlin quasielectrons
(QE) of the 2DEG to form fractionally charged excitons (FCX), $h$QE 
or $h$QE$_2$.
Different FCX states have different optical properties, and their 
stability depends critically on the presence of QE's in the 2DEG.
This explains discontinuities observed in the PL spectrum at such
(Laughlin) filling factors as $\nu={1\over3}$ or ${2\over3}$.
\end{abstract}
}

The photoluminescence (PL) spectrum of a quasi-two-dimensional 
electron gas (2DEG) measures the one-particle Green's function 
describing removal of an electron ($e$) at the position of the 
valence hole ($h$)\cite{heiman,pawel,kheng,x-pl,macdonald,rashba,%
chen}.
Thus, PL probes the electron correlations in the close vicinity 
of the (optically injected) hole, which, depending on the type of 
response of the 2DEG to the perturbation associated with this hole, 
may or may not resemble the original correlations of an unperturbed 
2DEG.
In an unperturbed 2DEG in a sufficiently high magnetic field $B$, 
the electrons fill a fraction $\nu$ of the degenerate lowest 
Landau level (LL).
The short range of the Coulomb $e$--$e$ repulsion in this LL causes
Laughlin correlations\cite{laughlin} (i.e., avoiding pair states 
of the smallest relative pair angular momentum\cite{parentage}) 
which lead to the incompressible-fluid Laughlin--Jain ground states
\cite{laughlin,jain} and the fractional quantum Hall (FQH) effect
\cite{tsui} at a series of fractional LL fillings, $\nu={1\over3}$, 
${2\over3}$, ${2\over5}$, etc.
Two effective parameters describe perturbation introduced by the 
hole: strength $U$ and range $D$ of its Coulomb potential $V_{UD}$.
Although the hole's density profile in the plane parallel to the 
2DEG is determined by its confinement to the lowest LL, potential 
$V_{UD}$ can still be varied by placing the hole at a finite distance 
$d$ from the 2DEG.
This is realized experimentally in asymmetric quantum wells or 
heterojunctions\cite{heiman}, where an electric field perpendicular 
to the 2DEG causes spatial displacement of electron and hole layers. 
Depending on the relation between $D$ and $U$ and the length and 
energy scales of the 2DEG (magnetic length $\lambda$ and the energy 
gap $\delta=\varepsilon_{\rm QE}+\varepsilon_{\rm QH}$ to create 
a QE--QH pair), two distinct types of response occur:

(i) In the ``strong-coupling'' regime, the hole binds one or two 
electrons to form neutral ($X$) or charged ($X^-$) excitons
\cite{kheng,x-pl,x-dot,x-fqhe}.
Although for different reasons (``hidden symmetry'' for $X$ 
\cite{dzyubenko} and Laughlin $e$--$X^-$ correlations for $X^-$
\cite{x-pl,x-fqhe}), both quasiparticles (QP) are weakly coupled to 
the remaining electrons, and PL probes {\em their} binding energy 
$\Delta$ and the optical properties (oscillator strength $\tau^{-1}$ 
and emission energy $\omega$) rather than the original correlations 
of the unperturbed 2DEG\cite{macdonald}.

(ii) In the ``weak-coupling'' regime, the distant hole interacts 
(weakly) with Laughlin excitations of the 2DEG (attracts quasielectrons 
QE and repels quasiholes QH), and the dynamics and PL of the system can 
be understood in terms of these three ($h$, QE, and QH) QP's
\cite{rashba,chen}.

In this Letter, we study the crossover between these two phases 
and identify another, ``intermediate-coupling'' regime.
In this regime, as in (ii), the Laughlin $e$--$e$ correlations 
dominate over the $e$--$h$ correlations, and neither $X$ nor 
$X^-$ states form.
However, in contrast to (ii), the $h$--QE attraction is sufficiently 
strong to cause binding of one or two QE's to a hole to form 
fractionally charged excitons (FCX) $h$QE or $h$QE$_2$.
The FCX's, together with a ``decoupled hole'' state $h$, are 
the relevant QP's whose interaction with one another and with 
the underlying Laughlin state is weak and can be treated 
perturbatively.
The single-particle properties of these QP's determine the 
low-energy dynamics and PL of the 2DEG at $d$ larger than but 
comparable to $\lambda$.
The critical dependence of the stability of different FCX's on 
the presence of QE's in the 2DEG causes discontinuities observed
\cite{heiman} in the PL spectrum at $\nu={1\over3}$ or ${2\over3}$, 
where the type of free Laughlin QP's changes (and the FQH effect
\cite{tsui} occurs).
These discontinuities resemble those found at $\nu=1$, 2, \dots\ 
\cite{pawel}, although the reconstruction of radiative QP's at 
integer $\nu$ is driven by the competition between $V_{UD}$ and 
the (single-particle) cyclotron energy, while the effect discussed 
here depends on the special form of many-body correlations in the 
lowest LL.
Let us also note that our results do not support the earlier theory
\cite{rashba} which involved neutral ``anyon excitons'' $h$QE$_3$ 
(we find that these complexes are neither stable nor radiative at 
any $d$ or $\nu$).

An infinite, spin-polarized 2DEG at a fractional filling factor 
$\nu$ is modeled by a finite system of $N$ electrons confined to the 
lowest shell (LL) of a Haldane sphere of radius $R$\cite{haldane}.
The angular momentum of this shell is $l=S$, half the strength of 
Dirac's magnetic monopole (defined in the units of elementary flux, 
so that $4\pi R^2B=2S\, hc/e$ and $\lambda=R/\sqrt{S}$).
The electrons interact with one another and with a valence hole 
of the same angular momentum $l_h=S$, moving in the lowest LL of 
a parallel 2D layer, separated from the 2DEG by a distance $d$.
After the constant cyclotron and Zeeman terms are omitted, the 
reduced $Ne$--$h$ Hamiltonian contains only the $e$--$e$ and 
$e$--$h$ interactions defined by a pair of 2D potentials, 
$V_{ee}(r)=e^2/r$ and $V_{eh}(r)=-e^2/\sqrt{r^2+d^2}$.

Using Lanczos-based algorithms we were able to diagonalize 
Hamiltonians of up to nine electrons and a hole at $2S\approx3(N-1)$
corresponding to $\nu\approx{1\over3}$.
Obtained energies and wavefunctions of low-lying $Ne$--$h$ states 
were labeled by total angular momentum $L$ and its projection $M$.
From the energy spectra, all bound states of a 2DEG coupled to a 
hole (depending on $d$ and $\nu$) were identified, and their binding 
energies $\Delta$ were calculated. 
From the wavefunctions, the PL energy $\omega$ and oscillator 
strength $\tau^{-1}$ of each QP were calculated.
The total QP angular momenta $l$ result from addition of angular 
momenta of their constituent particles ($h$, $e$, and QE).

Conservation of both orbital quantum numbers in a finite-size 
calculation is a major advantage of using Haldane's (spherical) 
geometry to model an infinite (planar) 2DEG with a full 2D 
translational invariance.
Conversion of the results between the two geometries follows from 
the exact mapping\cite{x-pl} between the two spherical numbers, 
$L$ and $M$, and two conserved quantities on a plane: total angular 
momentum projection ${\cal M}$ and an additional quantum number 
${\cal K}$ associated with partial decoupling of the center-of-mass 
motion in a homogeneous magnetic field\cite{avron}.
Resolution of both ${\cal M}$ and ${\cal K}$ (or $L$ and $M$) of 
the QP's is essential for identification of the optical selection 
rule: $\Delta{\cal M}=\Delta{\cal K}=0$ (or $\Delta L=\Delta M=0$) 
which results from the commutation of both ${\cal M}$ and ${\cal K}$ 
(or $L$ and $M$) with the PL operator\cite{x-pl}.

\begin{figure}[t]
\epsfxsize=3.40in
\epsffile{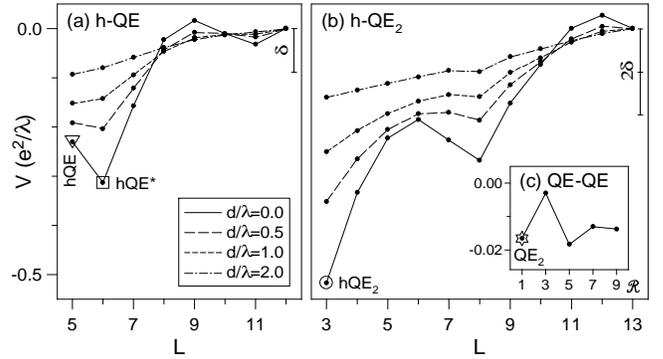}
\caption{
   Interaction pseudopotentials $V_{h-{\rm QE}}$ (a) and 
   $V_{h-{\rm QE}_2}$ (b) calculated in the $7e$--$h$ system, 
   and $V_{{\rm QE}-{\rm QE}}$ (c) calculated in the $11e$ system.
   $L$ is the pair angular momentum,
   ${\cal R}$ is the relative pair angular momentum,
   $\lambda$ is the magnetic length,
   and $\delta=\varepsilon_{\rm QE}+\varepsilon_{\rm QH}$ is 
   the Laughlin gap.
}
\label{fig1}
\end{figure}

The bound states of a hole and either electrons or QE's can be 
determined by studying the interactions of the constituent particles.
A two-body interaction is defined by a pseudopotential, the pair 
interaction energy $V$ as a function of an appropriate pair orbital 
quantum number.
The pseudopotentials most important for our analysis are shown in 
Fig.~\ref{fig1} for a few values of $d$.
Different bound states are marked on the curves for $d=0$.
The pseudopotentials of $h$--QE (a) and $h$--QE$_2$ (b) pairs were 
calculated in a $7e$--$h$ system at $2S=17$ and 16, respectively.
The allowed values of pair angular momentum $L$ result from addition 
of $l_h=S$ and either $l_{\rm QE}=S-N+2$ or $l_{{\rm QE}_2}=
2l_{\rm QE}-1$ (note that we use fermionic description of QE's
\cite{hierarchy}; $l_{\rm QE}$ is equal to angular momentum of 
the excited shell of Jain's composite fermions (CF)\cite{jain}).
To assure that exactly one or two QE's interact with a hole at any
$d$ (and no spontaneously created QE--QH pairs ``dress'' the FCX), 
the $e$--$h$ interaction used to calculate the $h$--QE$_n$ 
wavefunctions was multiplied by $\epsilon^{-1}\ll1$.
The pseudopotentials were calculated by subtracting from the eigenvalues 
obtained with this reduced potential the appropriate energy of the 2DEG 
together with the interaction energy of the hole and uniform density 
$\nu={1\over3}$ fluid, and then multiplying by $\epsilon$. 
Both pseudopotentials have been shifted so that $V_{h-{\rm QE}_n}=0$ 
at the maximum $L$ (maximum average $h$--QE$_n$ separation 
$\left<r\right>$).
In addition to the $h$QE and $h$QE$_2$ states, we also marked the 
$h$QE* state which later will be important in the analysis of PL.
The $V_{{\rm QE}-{\rm QE}}$ of an unperturbed 2DEG (c) was calculated 
for 11 electrons at $2S=28$.
For identical fermions, the pair quantum number increasing with 
increasing $\left<r\right>$ is the relative angular momentum 
${\cal R}=2l-L$ (an odd integer)\cite{parentage}.
Remarkably, despite QE's being charge excitations, the QE--QE 
interaction is not generally repulsive\cite{hierarchy}, and the low 
energy of the QE$_2$ molecule at ${\cal R}=1$ additionally stabilizes 
the $h$QE$_2$.

Occurrence of $h$QE$_n$ states of Fig.~\ref{fig1}(ab) requires the
presence of free or induced QE's in the 2DEG, and thus depends on 
both $\nu$ and the strength of $h$--QE$_n$ attraction relative to 
$\delta=\varepsilon_{\rm QE}+\varepsilon_{\rm QH}$ (marked in each 
frame).
At $d<2\lambda$, the $h$--QE attraction in both $h$QE and $h$QE* 
exceeds $\delta$, and these QP's will occur at any value of $\nu$ 
close to ${1\over3}$.
However, at $d>2\lambda$, the $h$--2DEG coupling is too weak to 
induce QE--QH pairs, the hole can only bind existing QE's, and 
$h$QE and $h$QE* will only occur at $\nu>{1\over3}$ (at $\nu<
{1\over3}$, holes repel free QH's and cause no local response 
of the 2DEG).
Similarly, the comparison of $V_{h-{\rm QE}_2}$ and $2\delta$ in 
Fig.~\ref{fig1}(b) shows that at sufficiently small $d$ the hole 
induces two QE--QH pairs, and the most stable QP at any value of 
$\nu$ close to ${1\over3}$ is $h$QE$_2$.

\begin{figure}[t]
\epsfxsize=3.40in
\epsffile{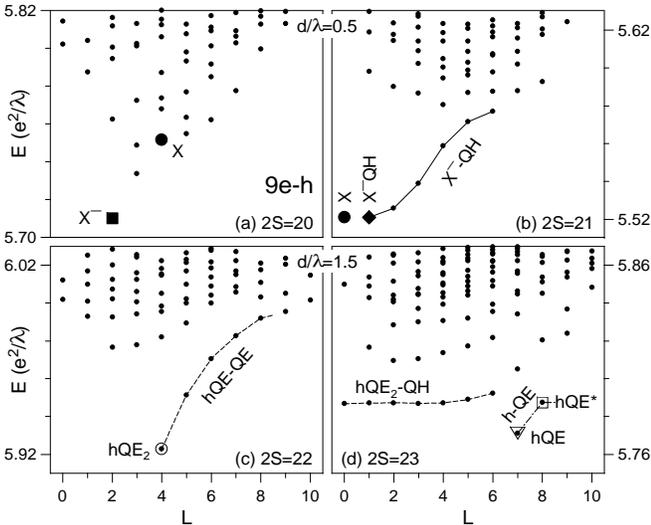}
\caption{
   Energy spectra (energy $E$ vs.\ angular momentum $L$) 
   of the $9e$--$h$ system on a Haldane sphere. 
   Symbols and lines mark different quasiparticles.
   $2S$ is the monopole strength,
   $d$ is the layer separation, and
   $\lambda$ is the magnetic length.}
\label{fig2}
\end{figure}

Knowing the QP's and their angular momenta allows understanding 
of the low-energy spectra of $Ne$--$h$ systems at any values of 
$d$ and $\nu$.
A few spectra for $N=9$ are shown in Fig.~\ref{fig2}.
At smaller $d$, the QP's are $e$, $X$, and $X^-$ with $l=S$, 0, 
and $S-1$, respectively.
The low-lying states describe either a Laughlin-correlated 
two-component $(N-2)e$--$X^-$ fluid or an $X$ (nearly) decoupled 
from a (also Laughlin-correlated) $(N-1)e$ state.
A system containing electrons and charged excitons can contain four 
types of Laughlin QP excitations: QE$_e$, QH$_e$, QE$_{X^-}$, or 
QH$_{X^-}$, whose angular momenta can be found from a generalized 
CF model\cite{x-fqhe}. 
In Fig.~\ref{fig2}(a), a standard CF-type analysis predicts that 
the ground state at $L=2$ contains a single QE$_{X^-}$.
In Fig.~\ref{fig2}(b), a band of QE$_{X^-}$--QH$_e$ pair states with
$1\le L\le6$ starts at the energy close to the lowest $(N-1)e$--$X$ 
state at $L=0$ ($X$ decoupled from the Laughlin $8e$ state).
Note that the present interpretation of Fig.~\ref{fig2}(b) 
invalidates the concept of a ``dressed'' exciton $X^*$\cite{apalkov} 
which turns out to be unstable toward the formation of an $X^-$ 
(coupling of an $X$ with $k\ne0$ to the 2DEG is too strong to be 
treated perturbatively).
At larger $d$, the QP's are $h$, $h$QE, and $h$QE$_2$ with
$l=S$, $N-2$, and ${1\over2}(N-1)$, respectively.
An isolated $h$QE$_2$ is the ground state at $L=4$ in Fig.~\ref{fig2}(c).
In Fig.~\ref{fig2}(d), two low-energy bands occur: at $L\le6$ the 
$h$QE$_2$ (weakly) interacts with a remaining QH, and the band at 
$L\ge7$ describes the $h$--QE dispersion of Fig.~\ref{fig1}(a), with 
$h$QE and $h$QE* being the two lowest states.

The recombination of QP's must {\em locally} conserve ${\cal M}$ and 
${\cal K}$.
The associated selection rules can be more easily obtained in the 
spherical geometry, where an area containing an isolated QP is 
represented by a whole (finite) system with appropriate $2S$, and 
the two conserved quantities are $L$ and $M$.
At smaller $d$, the only radiative state is $X$. 
The $X^-$ is dark because $l_{X^-}=S-1$ is different from $l=S$ of 
an electron left behind in the final state\cite{x-dot,x-fqhe}.
At larger $d$, efficient $h$QE$_n$ recombination must occur with 
the minimum number of QE's and QH's involved\cite{chen}.
For $\nu\approx{1\over3}$, such processes are $h{\rm QE}_n\rightarrow
(3-n){\rm QH}+\gamma$, where $n\le3$ and $\gamma$ denotes emitted 
photon.
To find the $h$QE$_n$ and $(3-n)$QH angular momenta we use $l_h=S$ 
and $l_{\rm QE}=l_{\rm QH}=S-N+2$, where $2S=3(N-1)-n$.
For the initial states, the result is: 
   $l_h={3\over2}(N-1)$, 
   $l_{h{\rm QE}}=N-2$,
   $l_{h{\rm QE}^*}=N-1$,
   $l_{h{\rm QE}_2}={1\over2}(N-1)$, and 
   $l_{h{\rm QE}_3}=3$.
For the final states:
   $L_{3{\rm QH}}\ge l_{{\rm QH}_3}={3\over2}(N-1)$,
   $L_{2{\rm QH}}=l_{{\rm QH}_2}-j$ 
      (where $l_{{\rm QH}_2}=N-1$ and $j$ is an even integer), and
   $l_{{\rm QH}}={1\over2}(N-1)$.
From the comparison of these values, it is clear that the only 
allowed radiative processes are:
\begin{eqnarray}
   h\phantom{{\rm QE}^*} &\rightarrow& {\rm QH}_3           +\gamma,
\nonumber\\
   h{\rm QE}^*           &\rightarrow& {\rm QH}_2           +\gamma,
\nonumber\\
   h{\rm QE}_2           &\rightarrow& {\rm QH}\phantom{_2} +\gamma.
\nonumber
\end{eqnarray}
Because $\tau^{-1}$ is proportional to the overlap between 
annihilated $e$ and $h$ charges, and $h$ is smaller than QH, 
the following ordering of PL oscillator strengths is expected:
$\tau_{h{\rm QE}_2}^{-1}>\tau_{h{\rm QE}^*}^{-1}>\tau_h^{-1}$.
Because of the high energy of the QH$_2$ and QH$_3$ molecules 
($V_{{\rm QH}-{\rm QH}}$ is strongly repulsive at ${\cal R}=1$
\cite{hierarchy}), we also predict:
$\omega_{h{\rm QE}_2}>\omega_{h{\rm QE}^*}>\omega_h$.

\begin{figure}[t]
\epsfxsize=3.40in
\epsffile{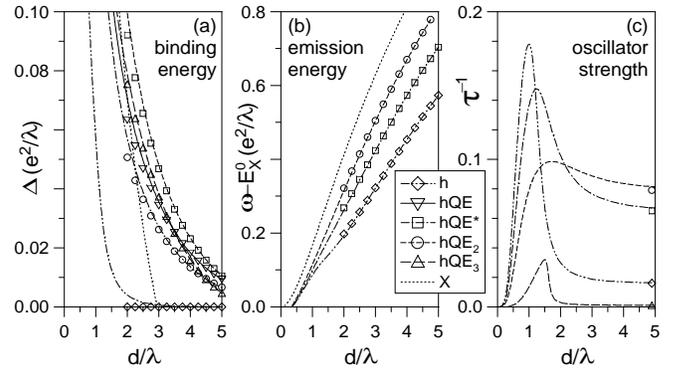}
\caption{
   Binding energy $\Delta$ (a), recombination energy $\omega$ (b), and 
   oscillator strength $\tau^{-1}$ (c) of different quasiparticles as 
   a function of layer separation $d$, calculated for the $8e$--$h$ 
   system.
   Lines (symbols) mark the exact (approximate; $\epsilon^{-1}\ll1$) 
   calculation (see text).
   $E_X^0$ is the exciton energy in the absence of the electron gas 
   and $\lambda$ is the magnetic length.}
\label{fig3}
\end{figure}

The values of $\Delta$, $\omega$, and $\tau^{-1}$, calculated from 
the exact wavefunctions of the $h$QE$_n$ states identified in the 
$8e$--$h$ spectra similar to those of Fig.~\ref{fig2}, are plotted 
as a function of $d$ in Fig.~\ref{fig3}.
$\Delta$ is the total interaction energy of the hole and either 
electrons or QE's, and does not include the attraction of the 
hole to the underlying Laughlin state.
The results of exact calculations shown in Fig.~\ref{fig3} confirm 
our predictions based on the analysis of the pseudopotentials of 
Fig.~\ref{fig1}.
In particular, $h$QE$_2$ is the most strongly bound and most strongly 
radiative complex in the entire range of $d$ in which the FCX's form. 
The change of correlations (replacing of $X$'s and $X^-$'s by 
$h$QE$_n$'s) between $d=\lambda$ and $2\lambda$ is best seen in the 
$\tau^{-1}$ curves in Fig.~\ref{fig3}(c).
Good agreement of the exact (lines) and approximate ($\epsilon^{-1}
\ll1$; symbols) calculation at $d>2\lambda$ indicates that the 
$h$QE$_n$'s are virtually uncoupled from the charge excitations 
of the 2DEG (and thus have well defined $n$ and $d$-independent 
wavefunctions).

Since the stability of different $h$QE$_n$ QP's depends critically 
on the presence of free QE's, their most striking signature in PL 
experiments should be discontinuities at Laughlin filling factors, 
such as $\nu={1\over3}$ or ${2\over3}$.
Schematic PL spectra at $\nu\approx{1\over3}$ predicted by our model 
are shown in Fig.~\ref{fig4}.
Consecutive frames correspond to increasing $d$ (although the actual 
critical values may change considerably in a more realistic model):
\begin{figure}[t]
\epsfxsize=3.40in
\epsffile{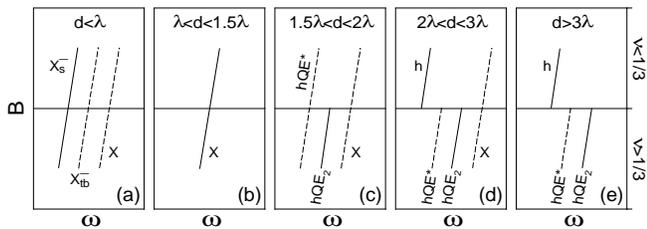}
\caption{
   Schematic PL spectra (recombination energy $\omega$ vs.\ 
   magnetic field $B$) at $\nu\approx{1\over3}$ and at different 
   layer separations $d$.
   Solid and dashed lines mark recombination from ground and 
   excited states, respectively.
   $\lambda$ is the magnetic length. 
}
\label{fig4}
\end{figure}
(a) The $X$ and {\em radiative} (not discussed here) $X^-$ states are 
observed\cite{x-pl}, the lowest being the bright spin-singlet $X^-$
(the singlet $X^-$ does not occur in the very high magnetic field 
limit, $\hbar\omega_c\gg e^2/\lambda$, emphasized in this paper). 
(b) The $X^-$'s unbind but the $X$'s still exist and dominate the 
PL spectrum (FCX's have smaller $\Delta$ and $\tau^{-1}$).
(c) The FCX's become visible due to their increasing $\Delta$; 
$h$QE's occur (and give rise to the $h$QE* recombination) at any
$\nu$ because of the spontaneous creation of one QE--QH pair in 
response to the hole charge; the (more strongly bound) $h$QE$_2$'s 
occur only at $\nu>{1\over3}$.
(d) Both $h$QE's and $h$QE$_2$'s exist only at $\nu>{1\over3}$; 
at $\nu<{1\over3}$, the hole causes no (local) response of the 2DEG, 
and the recombination occurs from the ``decoupled hole'' state $h$.
(e) The $X$'s unbind.

In conclusion, we have studied PL from a 2DEG in the FQH regime 
as a function of the $e$--$h$ layer separation $d$.
The QP's of this system have been identified, which (depending on 
$d$) consist of either one or two electrons or up to two Laughlin 
QE's bound to a hole.
The angular momenta, binding and recombination energies, and 
oscillator strengths of the QP's have been calculated and used 
to explain the numerical energy spectra and the experimental PL 
spectra of the 2DEG at an arbitrary $d$.
The discontinuities in the PL spectra of spatially separated 
systems at $\nu={1\over3}$ or ${2\over3}$ have been understood.

The authors acknowledge support by the Materials Research Program 
of Basic Energy Sciences, US Dept.\ of Energy and helpful discussions
with P. Hawrylak (NRC, Ottawa) and M. Potemski (HMFL, Grenoble).
AW acknowledges support from the KBN grant 2P03B05518.
\\[-6ex]

\end{document}